\documentclass[journal]{IEEEtran}

\usepackage{cite}
\usepackage{array} 
\usepackage{multirow}
\usepackage{amsmath,amssymb,amsfonts}
\usepackage{algorithm2e}
\usepackage{graphicx}
\usepackage{textcomp}
\usepackage{subcaption}
\usepackage{xcolor}
\usepackage{anyfontsize}
\usepackage[T1]{fontenc}
\usepackage{amssymb}
\def\BibTeX{{\rm B\kern-.05em{\sc i\kern-.025em b}\kern-.08em
    T\kern-.1667em\lower.7ex\hbox{E}\kern-.125emX}}
\usepackage{soul}

\begin{document}

\title{Testing the Usability and Accessibility of Smart TV Applications Using an Automated Model-based Approach}

\author{Miroslav Bures, Miroslav Macik, Bestoun S. Ahmed, Vaclav Rechtberger, and Pavel Slavik
\thanks{This study is conducted as a part of the project TACR TH02010296 ``Quality Assurance for Internet of Things Technology''. This work has been supported by the OP VVV funded project CZ.02.1.01/0.0./0.0./16\_019/0000765 „Research Center for Informatics“. \textit{(Corresponding author: B. Ahmed)}}

\thanks{M. Bures and V. Rechtberger are with the Department of Computer Science, Faculty of Electrical Engineering, Czech Technical University, Karlovo nam. 13, Prague, Czech Republic, email: buresm3@fel.cvut.cz}

\thanks{M. Macik and P. Slavik are with the Department of Computer Graphics and Interaction, FEE, Czech Technical University in Prague, Karlovo nam. 13, Prague, Czech Republic, email: macikmir@fel.cvut.cz, slavik@fel.cvut.cz}

\thanks{B. Ahmed is with the Department of Mathematics and Computer Science, Karlstad University, Sweden and the Department of Computer Science, Czech Technical University, Karlovo nam. 13, Prague, Czech Republic,
email: bestoun@kau.se}%

}

\maketitle

\begin{abstract}
As the popularity of Smart Televisions (TVs) and interactive Smart TV applications (apps) has recently grown, the usability of these apps has become an important quality characteristic. Previous studies examined Smart TV apps from a usability perspective. However, these methods are mainly manual, and the potential of automated model-based testing methods for usability testing purposes has not yet been fully explored. In this paper, we propose an approach to test the usability of Smart TV apps based on the automated generation of a Smart TV user interaction model from an existing app by a specialized automated crawler. By means of this model, defined user tasks in the Smart TV app can be evaluated automatically in terms of their feasibility and estimated user effort, which reflects the usability of the analyzed app. This analysis can be applied in the context of regular users and users with various specific needs. The findings from this model-based automated analysis approach can be used to optimize the user interface of a Smart TV app to increase its usability, accessibility, and quality.
\end{abstract}

\begin{IEEEkeywords}
Usability Testing, Model-based Testing, User Interface Quality, Smart TV application.
\end{IEEEkeywords}

\IEEEpeerreviewmaketitle

\section{Introduction}

Currently, Smart TVs are coming to dominate the television market, and the number of connected TVs is growing exponentially. This growth is accompanied by an increase in consumers and the use of Smart TV apps that drive these devices. Smart TV apps fully interact with the user via a visualized UI and a remote device. Due to the increasing demand for Smart TV apps, especially with the rise of the Internet of Things (IoT), developing new usability testing methods for these apps is essential. The classic User Interface (UI) evaluation approaches for usability testing are based on mainly manually performed testing with respect to the UI of the System Under Test (SUT) \cite{dumas1999practical}. The potential of automated generation of a UI model from an existing Smart TV app combined with model-checking principles has not been fully explored.

To this end, the motivation of this study is threefold. First, the combination of UI model generation from an existing Smart TV app with model-checking principles to detect possible UI design suboptimality is not sufficiently covered in the literature. Second, concerns related to the usability of Smart TV apps were raised by Ingrosso et al. \cite{ingrosso2015ux} and Alam et al. \cite{alam2017review} in 2015 and 2017, respectively. In fact, Alam et al. \cite{alam2017review} discussed a number of potential usability issues of Smart TV apps. Considering the growth of the Smart TV market and the increase in app users, the focus on the usability testing of these apps must be intensified to prevent the usability problems reported by previous studies \cite{ingrosso2015ux, alam2017review}. A systematic and efficient usability testing method for Smart TV apps should be provided. Third, the current UI testing studies focus on various devices and types of apps. The Smart TV app domain remains relatively underexplored by relevant studies.

Based on the motivations mentioned above, the objective of this paper is to propose and verify an automated model-based method to detect possible design flaws or suboptimalities in the UI of a Smart TV app. We propose a method based on the analysis of the UI model of a Smart TV app that is acquired automatically by a specialized crawler. Defined user tasks in the Smart TV app are mapped to this model and then evaluated by a set of rules to verify feasibility and effectiveness of these tasks in which the user interacts with the app's UI. The context of the user interacting with an app is reflected in these rules. This context is expressed by a set of configuration constants, i.e., user capability to perform individual actions in the UI, device factor, environmental factor, and a default user effort of the individual actions in the UI. In this context, we can model users with various specific needs. The verification rules assess the feasibility of the task in the app for the user in a particular context and estimate the length to detect potential suboptimalities in the UI design or to detect repetitive steps in the UI needed to achieve the task. The findings of this analysis can help UI designers and app developers to optimize their UI in consideration of both the specific features of the Smart TV app and the particular needs of a user. This method can also aid the evaluation of user feedback on the quality of the app's UI in an independent objective manner. The contributions of this paper can be summarized as follows:

\begin{itemize}
\item We present an approach that potentially synergizes usability testing and functional testing based on the underlying model-based testing principles.

\item We propose an innovative method that enables analysis of the feasibility and ease of user tasks in a UI and assessment of the optimality based on a UI model that is generated automatically by a special crawler. Thus, an up-to-date and accurate design model of the UI from the design phase of the project is not needed.

\item We propose a novel application of model-based UI analysis in the Smart TV domain, which has not been sufficiently explored.

\item We report the parametrization of the user interaction model for Smart TV apps that is calibrated during several sets of experiments performed with real users.

\end{itemize}



\section{Related Work}
\label{chap:related_work}

Smart TV represent prospective stream of consumer electronic development. Compared to traditional TV, besides the possibility to personalize their user environment \cite{kim2016personalized}, users appreciate variety of applications that can be installed in the smart TV set, spanning from various games, media and infotainment applications to various services, including employment of smart TV sets in various home IoT solutions. Especially this field is a subject of recent research and development, for instance controlling of smart home appliances  \cite{Kim2013,Kim2015}, smart light management system \cite{Chun2013} or the whole smart home solution \cite{Cabrer2006,jalal2018internet} using a smart TV set, various personal healthcare application employing smart TV, for instance \cite{Vavilov2014} or personal sleep management employing a video analysis using a smart TV application \cite{Fan2014}. As another example, smart home security system using cameras and smart TV set can be given  \cite{Erkan2015}. Integration of smart TV sets into various smart home systems and services as well as increasing popularity of smart TV among users also increase requirements on usability of their applications.

Regarding usability testing of smart TV applications, previous work related to manual usability testing and assessments can be found. To give few examples, Shin et al. \cite{shin2013smart} examined the users' attitude and perception of Smart TV devices from a usability perspective.

Ingrosso et al. \cite{ingrosso2015ux} examined the usability of Smart TV apps using a case study of a T-commerce application. 

A number of potential usability issues of Smart TV apps were discussed in a more recent analysis by Alam et al. \cite{alam2017review}. These recent studies can also be seen as motivation to develop specific usability testing methods to improve the general usability of Smart TV apps.

Regarding the automation of usability tests, several previous projects can be identified. For example, automated testing of usability and accessibility of web pages has been proposed by Okada et al. \cite{okada2008automated}. Here, the proposed system collects logs from users' interaction with the SUT. The usability and accessibility were evaluated by comparing the logs with hypothetical ideal scenarios. Also, more formal approaches to usability testing have been examined in the literature to enable a more systematic approach to the design of the test automation system. Gimblett and Thimbleby \cite{gimblett2013applying} proposed a testing approach using a theorem discovery method to find and check usability heuristics automatically. Here, sequences of equivalent or very similar user inputs and their effect on the SUT were analyzed \cite{gimblett2013applying}. 

Cassino and Tucci \cite{cassino2011developing} proposed an approach to evaluate the interactive visual environments, which is based on SR-Action Grammars \cite{cassino2003sr}. This approach aids developers to create applications in which the UI respect defined usability rules. The practical implementation of the method resulted in the automatic usability verification tool \cite{cassino2011developing}. The formal specification is created from the SUT and used for subsequent usability checks and as particular usability rules, set of Nielsen heuristics \cite{nielsen2005ten} were employed in the proposed tool.

However, during our analysis of the state of the art, we have found only a few studies related to the automated usability testing of Smart TV apps based on a model created by an automated scan of the app's UI. Previous effort regarding the modeling of the smart TV app has been done by Cui et al. \cite{Cui2017}. Instead of a user interaction model with the app's UI as we propose, Cui et al. employed the hierarchical state transition matrix (HSTM), which is based on a state machine and hierarchical structure of the app.

Several crawlers creating a model for the UI have been presented in the literature for mobile and web apps. For instance, the projects by Mesbah et al. \cite{Mesbah2012crawling} for web applications, Memon et al. \cite{memon2003gui} for thick-client app UIs, Amalfitano et al. \cite{amalfitano2011gui,Amalfitano2015MobiGuitar}, and Wang et al. \cite{wang2014automatic} for mobile apps. Also, the universal frameworks allowing connection to a particular app's UI by a modular interface, as proposed by Nguyen et al. \cite{nguyen2014guitar}.

The concepts presented in this paper can also be conceptually compared to the model-checking approach. However, the applications of model-checking techniques usually focus on the detection of potential functional defects on various levels of the SUT in its classical form \cite{berard2013systems}, or when model-checking is combined with dynamic testing \cite{Godefroid2018combining}. As modeling structures, different formal notations and employed currently. These notations include finite state machines and their various extensions and modifications for the modeling of discrete systems \cite{Seshia2018modeling}, Petri nets or marked graphs for the modeling of concurrent processes, or hybrid automata or real-time temporal logics to model real-time systems \cite{Seshia2018modeling}.

Using the model-checking approach for UI usability testing is relatively under-explored in the literature. Harrison et al. \cite{harrison2017verification} focused on this domain recently, using temporal logic as an underlying model of the SUT.

\section{Overview of Our Proposed Approach}
\label{chap:overview_of_the_method}

The proposed method is applicable mainly to Smart TV apps during the development and testing process. However, the method can be applied to apps in alpha and beta testing or even production run, when the users report UI suboptimalities during their interaction with the app. Different types of suboptimalities exist, such as (1) user discomfort, (2) confusing organization of the individual elements of the UI, (3) too long or confusing sequence of steps to be taken to achieve frequently performed tasks, and (4) suboptimality of the app's UI for users with specific needs of particular category, or any other UI design flaws.

These suboptimalities are detected by metrics based on the proposed user interaction model (defined in Section \ref{subsec:user_interaction_model}) and the execution time of user scenarios.

The following steps summarize the conceptual process of the proposed approach:  

\begin{itemize}
\item The UI of the app is scanned by a special crawler (described in detail in Section \ref{sec:automated_model_creation_from_SUT}) that creates an extensive user interaction model of the Smart TV app (described in Section \ref{subsec:user_interaction_model}).

\item The user (the UI designer or the developer) defines a set of test scenarios. The scenarios capture the most frequent user tasks to be performed in the app and/or the user tasks that are reported as problematic from a usability/accessibility viewpoint by users or usability testers of the app.

\item Defined test scenarios are captured in the user interaction model using the specialized Model-based Testing (MBT) platform (details follow in Section \ref{sec:implementation}).
  
\item The context in which the defined test scenarios are assessed is defined using a set of configuration constants (discussed further in Section \ref{subsec:model_parametrization}).

\item A set of verifications is performed for each of the scenarios and defined context. These verifications include feasibility assessment of the scenario in the app's UI, user effort needed to execute the scenario and repetition of IU elements. The exact description of these verifications is presented in Section \ref{sec:ui_analysis}.

\item During the removal of the UI design problems identified in the previous step, the UI designer edits the SUT user interaction model in the MBT platform (more possible transitions or shortcuts in the SUT UI can be added, for instance). After these corrections, scenarios that were evaluated as problematic during the previous step can be reanalyzed until satisfactory results are achieved.

\item Finally, the adjustments in the user interaction model can be transformed into a set of change requests for the UI development team.

\end{itemize}

The used MBT system\footnote{http://still.felk.cvut.cz/oxygen/} is an experimental platform for process and path-based testing developed and issued by the Software Testing IntelLigent Lab (STILL), Dept. of Computer Science, FEE, Czech Technical University in Prague. The application supports creation of user models via a graphical UI and employs a set of algorithms to validate the created models and generate test cases from these models.


\section{User Interaction Model}
\label{chap:model}

Our proposed approach is based on the user interaction model (explained in Section \ref{subsec:user_interaction_model}) and its parametrization that reflects the context. The suggested values for the Smart TV domain are discussed in Section \ref{subsec:model_parametrization}.

\subsection{Model Definition}
\label{subsec:user_interaction_model}

A user's interaction with the Smart TV app's UI is abstracted as the user interaction model. Here, we use a directed multigraph to describe the model as $G=(N,E,n_{s},N_{e},s,t)$, such that $N\neq\emptyset$ is a finite set of nodes, $E$ is a set of edges, $s:E\rightarrow N$ assigns each edge to its source node and $t:E\rightarrow N$ assigns each edge to its target node. The node $n_{s}\in N$ is the initial/start node of the graph $G$, and $N_{e}=$ \{$n_{e}\mid n_{e}\in N$ has no outgoing edge \} defines nonempty set of end nodes of graph $G$. A node in the directed graph models a screen or a screen element of the UI. A screen element is a standalone clickable part of the screen layout or a nested container on the screen.

A graph edge in the model represents a transition between nodes via the interactive (control) element. Each transition $e \in E$ can be triggered by an \textbf{input action} $a(e)$. An input action is a physical action of the user on the remote control device that leads to transition $e$ in the app. Consider the remote control device as an example for a Smart TV app. Here, the input actions are events sent from the device to the Smart TV app when a user presses UP, DOWN, OK, or another button. An edge $e$ can have identical source and target node, making a simple loop; this case models the situation where an input action $a(e)$ does not trigger a transition between nodes on the app's UI but changes an internal state of the app.

The Tested User Scenario $t$ is an ordered sequence of nodes $N_t \subseteq N$ and edges $E_t \subseteq E$ which have to be visited during the execution of the user scenario. The $n_{1} \in N_t$ is a starting node of $t$ and $n_{n} \in N_t$ is a terminal node of $t$. A set $T$ is a set of all Tested User Scenarios. The nodes and edges in $t$ can repeat.

User scenario path $p(t)$ of the tested user scenario $t$ is a path in $G$ that contains the nodes $N_t$ and the edges $E_t$ of $t$ and can also contain other nodes or edges of $G$. $p(t)$ starts with $n_{1}$ and ends with $n_{n}$. The order of $N_t$ and $E_t$, as defined in $t$, is maintained in $p(t)$. Furthermore, $|p(t)|$ denotes the number of edges of $p(t)$, and $nodes(p(t))$ denotes the unique number of nodes of $p(t)$. Note that $t$ itself is not necessarily a path of $G$. Additionally, as the nodes and edges in $t$ can repeat, $p(t)$ is not necessarily the shortest path from $n_s$ to a node from $N_e$.

$C$ is the context in which the user accesses the app's UI. The user effort required to perform a transition $e \in E$ is

$\mathcal{E}(e,C) = \delta(a(e)) \times \frac{1}{UC(a(e),C)} \times \frac{1}{\mathcal{E}_{dev}(C)} \times \frac{1}{\mathcal{E}_{env}(C)}$, 

and the total user effort of user scenario path $p(t)$ is  

${\displaystyle \mathcal{E} (p(t),C)=\sum _{i=1}^{|p(t)|}\mathcal{E}(e_{i},C)}, e_i \in p(t)$,

where $UC(a(e),C)$ is the user capability to perform an action $a(e)$ in context $C$ (0 - user is unable to perform the action, 1 - user is able to perform the action with standard effort). $\mathcal{E}_{dev}(C)$ is the device factor, and $\mathcal{E}_{env}(C)$ is the environmental factor. $\delta(a(e))$ is the default effort of the particular action measured in milliseconds, including the time of cognitive effort to operate and the time to interact with the UI. The other constants, $UC$, $\mathcal{E}_{dev}$, and $\mathcal{E}_{env}$, are unitless.

The suggested values of $UC$, $\mathcal{E}_{dev}$, $\mathcal{E}_{env}$, and $\delta$ are discussed further in Sections \ref{subsec:model_parametrization} (initial values of the constants) and \ref{subsec:experiment_results} (refined values of the constants after the experiments). The total user effort is further used in the assessment of defined tested scenarios $T$ in the UI modeled by $G$ (details are provided in Section \ref{sec:ui_analysis}).



\subsection{Model Illustration}

\begin{figure}
\centering
  \includegraphics[width=0.7\columnwidth] {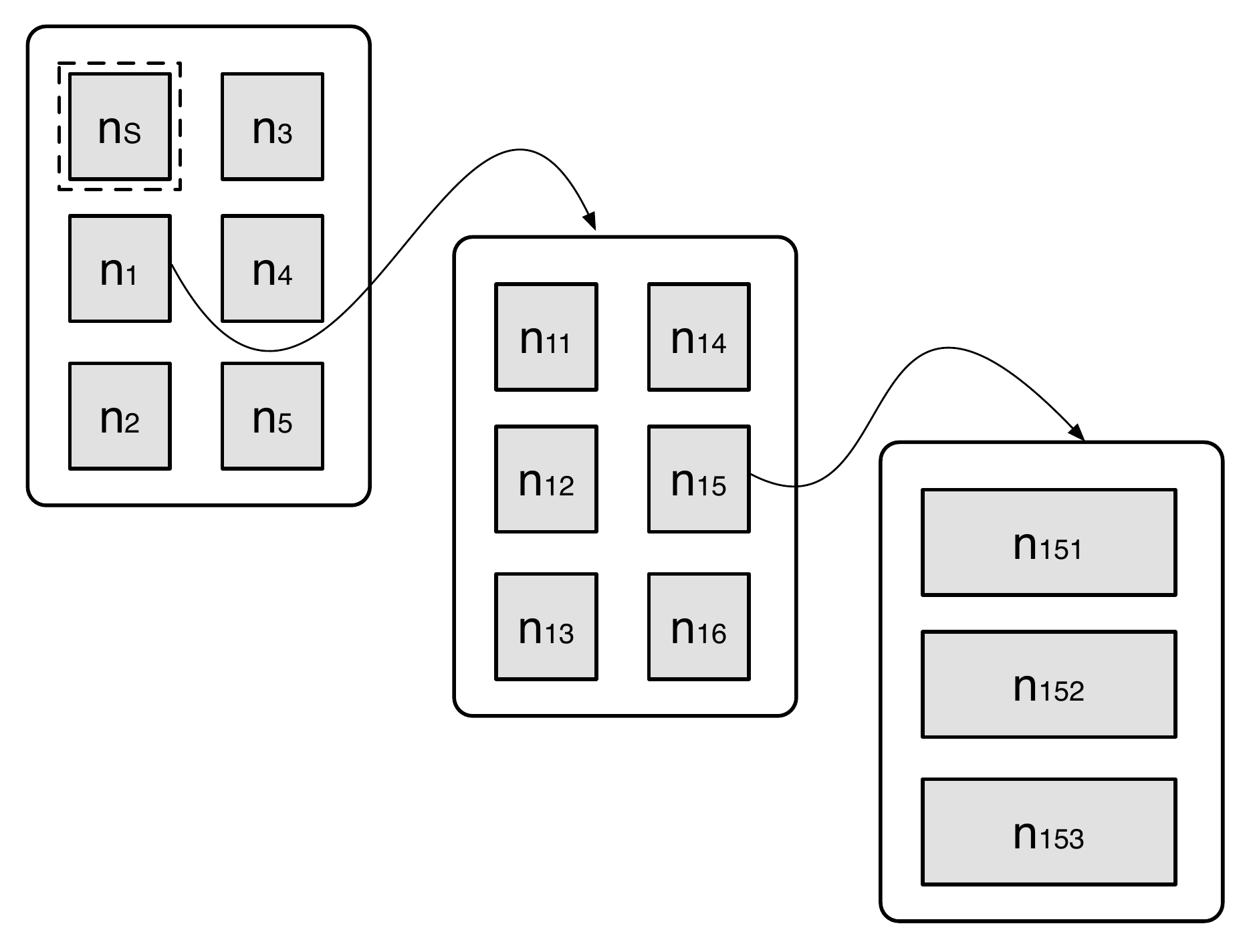}
  \caption{An abstracted example of the Smart TV app's UI}
  \label{fig:example_abstracted_UI}
\end{figure}

In this section, we demonstrate the user interaction model concepts using an abstracted example. Figure \ref{fig:example_abstracted_UI} shows three screens of a sample Smart TV app that contain various screen elements ($N$). Element $n_s$ is an initial screen element of the app. Using the remote control device, the user triggers possible transitions in the UI ($E$), and his focus changes to another screen element during this process.

\begin{figure}
\centering
  \includegraphics[width=0.7\columnwidth] {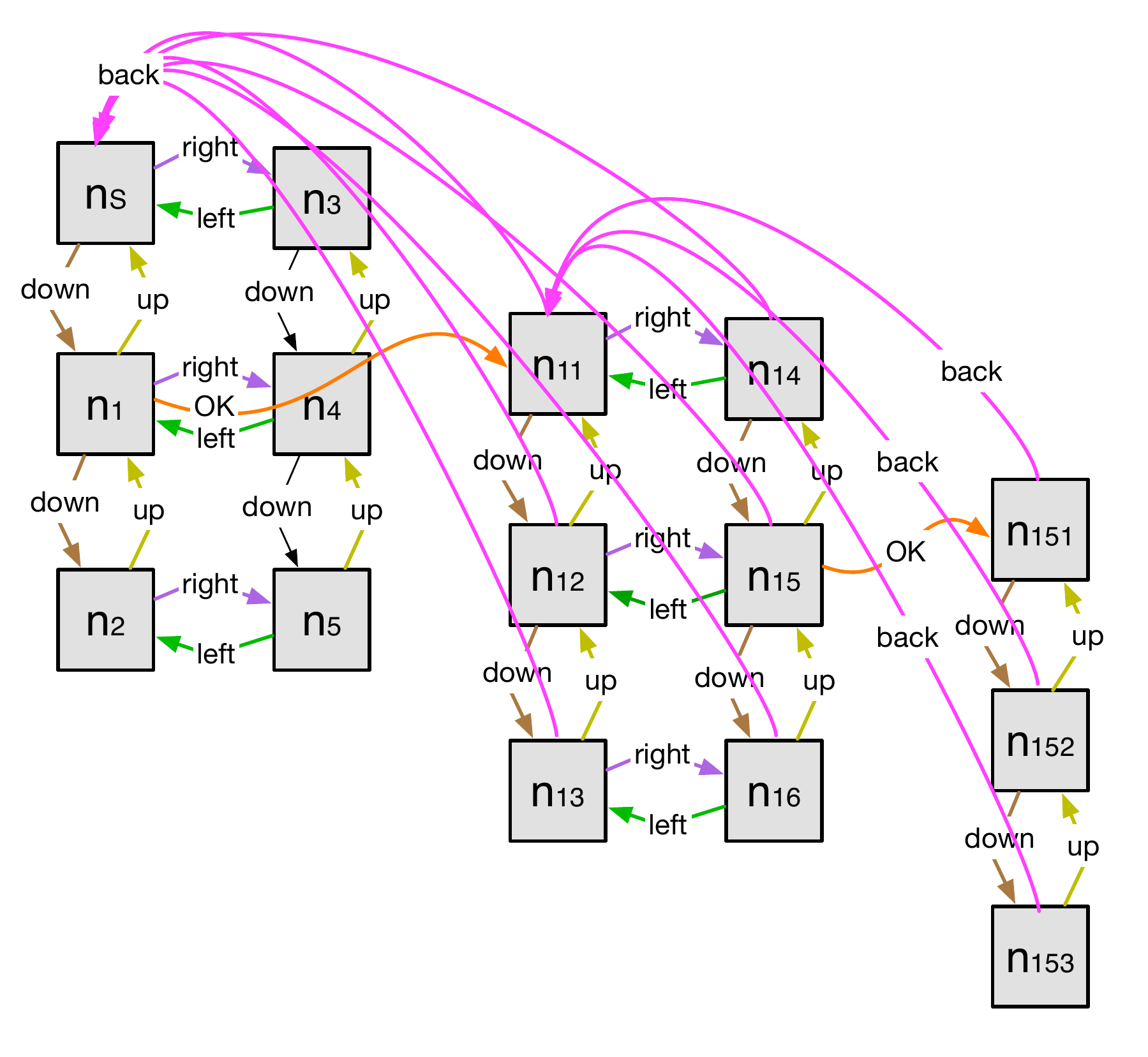}
  \caption{Sample user interaction model created for the example}
  \label{fig:sample_model}
\end{figure}

All possible paths that can be taken in this example are depicted in Figure \ref{fig:sample_model}; this is also the model that will be produced by the specialized crawler used in the proposed approach (a detailed description follows in Section~\ref{sec:automated_model_creation_from_SUT}). The outcome of this crawling process is a directed graph generated to model the elements of the app. 

\subsection{Parametrization of the Model}
\label{subsec:model_parametrization}

User effort depends on mainly the contextual circumstances, which we model by the context $C$. As an example, we take a system (i.e., a Smart TV app) that is controlled by a person challenged with a serious dexterity issue (i.e., quadriplegic). This person controls the app with a special controller that allows six actions (left, right, up, down, back, OK). Performing the individual actions with the controller requires different effort from the user and allows different efficiency. The basic actions -- left, right, back, OK -- are easy to perform. By contrast, significantly more effort is required to perform the remaining two actions (i.e., up and down). Hence, in different contexts, the settings of $UC$, $\mathcal{E}_{dev}$ and $\mathcal{E}_{env}$ would logically be different. Thus, we need to perform an initial setting of these constants, including $\delta$. Additionally, we need to calibrate these constants in the experiments. In this paper, we used six main actions by which the user can interact with the app's UI. Those actions are represented by the buttons UP, DOWN, LEFT, RIGHT, OK and BACK on the remote control device.

As a baseline, we consider the context $C_s$, which models a user without any special needs or disabilities. We also consider a standard Smart TV set with no environmental factors that might make the interaction with the Smart TV set more difficult. Table \ref{tab:initial_parametrization} shows the first setting of $\delta$, $UC$, $\mathcal{E}_{dev}$ and $\mathcal{E}_{env}$, or $C_s$, based on our previous empirical investigations. As $\delta$ aggregates the time of the user's cognitive preparation to perform an action in the app and the time needed to interact with the UI by the respective remote control button, the value of $\delta$ is higher for actions such as OK and BACK. After the pressing OK or BACK button, the user moves to a new UI screen, which must be analyzed before taking the next action to complete a task. Hence, the time needed for cognitive preparation is longer.

\begin{table}
\centering
\caption{An initial model parametrization} 
\scriptsize
\begin{tabular}{|l|c|c|c|c|}
\hline
action & \multicolumn{1}{l|}{$\delta(a(e))$} & \multicolumn{1}{l|}{$UC(a(e),C)$} & \multicolumn{1}{l|}{$\mathcal{E}_{dev}(C)$} & \multicolumn{1}{l|}{$\mathcal{E}_{env}(C)$} \\ \hline
LEFT   & 800                              & 1.0                              & \multirow{6}{*}{1.0}               & \multirow{6}{*}{1.0}                    \\ \cline{1-3}
RIGHT  & 800                              & 1.0                              &                                    &                                         \\ \cline{1-3}
UP     & 800                              & 1.0                              &                                    &                                         \\ \cline{1-3}
DOWN   & 800                              & 1.0                              &                                    &                                         \\ \cline{1-3}
OK     & 2500                              & 1.0                              &                                    &                                         \\ \cline{1-3}
BACK   & 1500                              & 1.0                              &                                    &                                         \\ \hline
\end{tabular}
  \label{tab:initial_parametrization}
\end{table}


\section{Automated Model Creation from the Smart TV App}

\label{sec:automated_model_creation_from_SUT}

The user interaction model $G$ introduced in Section \ref{subsec:user_interaction_model} is created by a specialized crawler that we developed for this purpose. The crawler starts at a defined screen $n_s$ of the Smart TV and explores its screens. During this process, only the code of the app screen is analyzed, and no knowledge of the internal structure of the app's code is obtained. On each screen, the crawler analyzes the available nested containers by examining each clickable element. During this analysis, each clickable screen or individual nested container is assigned a separate node in the $G$, being dynamically constructed during the crawling. 
The exploration process stops when no more clickable element is available to be explored or when a defined termination criterion has been met. The termination criteria are defined by a number of nodes $|N|$ in the created mode $G$. The termination criteria are used for dynamically generated UIs of apps with an online content (which essentially create an infinite space to explore). When the exploration process terminates, $N$ contains all the examined nodes.

When the crawler arrives at a screen or screen nested container, it examines the user actions available on this screen. This is done by simulating the user's remote control by pressing UP, DOWN, LEFT, RIGHT, OK and BACK buttons. Identified possible actions leading to a transition to next screens or screen nested containers are then added as edges to $G$.

When the crawler finishes the exploration of the SUT UI, $E$ contains all the possible transitions available from the screens and screen nested containers contained in $N$. The set $N_e$ contains all screens (or screen nested containers) for which no outgoing action is available (in the case of a well-designed Smart TV app, $N_e$ should be empty).

Regarding the time requirements to create the user interaction model $G$ via the crawler, the initial configuration of the crawler for a new Smart TV app takes up to 30 minutes for a completely new user. The crawling process itself depends on the size of the explored space; however, the run time of the crawler did not exceed 60 minutes for the testing app used in this study.

\section{Automated UI Analysis of the Smart TV App}
\label{sec:ui_analysis}

In our approach, we use the user interaction model $G$ that is generated by our automated crawler. As mentioned in Section \ref{subsec:user_interaction_model}, we designed the crawler to scan the Smart TV app and create the model without knowing the internal structure of the app code. The details of the crawler implementation and the full source code are available for download\footnote{Smart TV crawler download page https://github.com/bestoun/EvoCreeper}. A step-by-step running example of the crawler can be found in \cite{BestounSmartTVReview2019}. The following points detail the concepts of our approach:

\begin{enumerate}
\item The UI of the Smart TV app is scanned by our crawler, which creates model $G$ as the output.

\item A set of tested user scenarios $T$ is defined by a tester using. The scenarios include the most frequent user tasks in the app and/or the user tasks in the app that are reported as problematic steps from a usability/accessibility perspective.

\item Each $t \in T$ is mapped to the nodes and edges of $G$ in the MBT framework (details follow in Section \ref{sec:implementation}).
  
\item The user context $C$ in which the defined tested user scenarios $T$ will be assessed is defined. Namely, the values of $UC$, $\mathcal{E}_{dev}$, $\mathcal{E}_{env}$ and $\delta$ are set for the individual actions that can be invoked by the remote control.

\item The following set of verifications is performed for each $t \in T$:
\begin{enumerate}

 \item User scenario path $p(t)$ is constructed for $t$. If $p(t)$ does not exist, this fact indicates a UI design flaw. If this check is passed, then perform the following checks:

    \item If $p(t)$ is not a simple path, compute the node repetition 

  ${\displaystyle nr(p(t))={\frac {|p(t)|+1}{nodes(p(t))}}}$. Then, $nr(p(t))>nr_{threshold}$ may indicate possible UI design suboptimality. $nr_{threshold}$ is discussed in Section \ref{subsec:initial_values_of_thresholds}. 
  
\item $|p(t)|>|p(t)|_{threshold}$ may indicate possible UI design suboptimality. $|p(t)|_{threshold}$ is discussed in Section \ref{subsec:initial_values_of_thresholds}.

\item $\mathcal{E} (p(t),C)$ is computed:

\begin{enumerate}

  \item $\mathcal{E} (p(t),C) = \infty$ (or division by 0) indicates that $p(t)$ is infeasible for a particular user in context $C$ (typically, a limit for a user with a specific need).
  
  \item 
  $\mathcal{E} (p(t),C) > \mathcal{E}_{threshold}$
  may indicate possible UI design suboptimality.
  $\mathcal{E}_{threshold}$ is discussed in Section \ref{subsec:initial_values_of_thresholds}. 

\end{enumerate}
   
\end{enumerate}

\item To remove the UI design problems identified in the previous steps, the UI designer can edit $G$ in the MBT environment by adding an edge (or a set of edges), adding a node (or a set of nodes), or generally updating the model. Then, the problematic scenarios can be reanalyzed (repeat step 5) until the defined verification rules are satisfied.

\item The adjustments in $G$ can be transformed to a set of change requests for the UI development team to repair the detected problems or suboptimalities in the Smart TV app.

\end{enumerate}

When step 7 results in a change in the UI, steps 1-7 can be repeated to verify the suitability of the changes from the usability perspective. The whole cycle of steps 1-7 can be repeated several times until the optimal result is achieved.

\subsection{Initial Values of Thresholds}\label{subsec:initial_values_of_thresholds}

For the verification rules defined in Section \ref{sec:ui_analysis}, step 5, we set the following initial values of the thresholds. We set the value of $nr_{threshold}$ to 1.5, the value of $|p(t)|_{threshold}$ to 20 and the value of $\mathcal{E}_{threshold}$ to 25000. These values are based on our previous experience, and they are further adjusted based on feedback from the experiments in Section \ref{subsec:experiment_results}.

\section{Experimental Verification}
\label{chap:experimental_verification}

We have verified our proposed approach in an experimental evaluation study consisting of the technical verification of the methods and experiment with a group of Smart TV users. The following sections detail the experimental procedures and the evaluation results.

\subsection{Experiment Method}
\label{subsec:experiment_method}

The experiments were conducted in a sequence of the following steps:

\begin{enumerate}

\item We selected an open source Smart TV app\footnote{https://github.com/daliife/Cinemup} (further referred as testing app) as an SUT to be analyzed by our specialized crawler to create user interaction model $G$. 
  
\item \label{item.testbed_setup} We configured a special testbed setup for the experiments that consisted of Smart TV environment web simulator with an installed testing app with a special logging mechanism to capture user actions. In addition, this mechanism counts the exact time at which the user executes a~particular action (represented by an edge or node of $G$) on the app and the remote control button that triggered the action.

\item We defined a set of four tested user scenarios $T$: one to be used as a training scenario for the experiment participants and three to collect the experimental data. The scenarios were deliberately defined in less detail (capturing only a generally defined user task, not a sequence of main screens and input actions to be visited/achieved on the app). The user scenarios used in the experiment are described in detail in Section \ref{subsec:user_scenarios}.
  
\item As described in the method defined in Section \ref{sec:ui_analysis}, each $t \in T$ is mapped to the nodes and edges of $G$ in the MBT environment.

\item \label{item.run_verifications} For each $t \in T$, we ran the set of verification procedures defined in Section \ref{sec:ui_analysis}. We implemented the verification rules as a part of the MBT platform. The initial configurations of $UC$, $\mathcal{E}_{dev}$, $\mathcal{E}_{env}$ and $\delta$ are presented in Section \ref{subsec:model_parametrization}.

\item \label{item.users_test} Concurrently, each $t \in T$ was implemented in the testbed by 25 independent users recruited from the students of a software testing course. The users were instructed to perform the user scenario as specified by $t$. The logging mechanism logged their activities on the app.
    
\item \label{item.compare_resutls} We compared the results obtained from the application of the verification rules (step \ref{item.run_verifications}) and the independent test by users (step \ref{item.users_test}). Namely, we compared the total time needed to accomplish the user scenarios and the length of the user scenario paths on the UI measured as the number of transitions, and we analyzed the length of individual user scenario paths invoked on the UI by the remote control buttons. The results are presented in Section \ref{subsec:experiment_results}.
  
\item \label{item.users_test_2} We repeated step \ref{item.users_test} again with another group of 24 participants. The details of the second verification are in Section \ref{subsec:experiment_results}. 
  
\item \label{item.adjust_configuration} Based on feedback from the comparison results and from the second experiment, we adjusted the configurations of $UC$, $\mathcal{E}_{dev}$, $\mathcal{E}_{env}$ and $\delta$. Additionally, we adjusted the values of the thresholds $nr_{threshold}$, $|p(t)|_{threshold}$ and $\mathcal{E}_{threshold}$. We present the updated values in Section \ref{subsec:experiment_results}.
    
\item \label{item.run_verifications_2} We repeated step \ref{item.run_verifications} with the adjusted configurations of $UC$, $\mathcal{E}_{dev}$, $\mathcal{E}_{env}$ and $\delta$.
    
\item Again, we compared the results obtained from the verification rules on the app and the independent test by the users to check for improvement in the method configuration. The details of this second verification follow in Section \ref{subsec:experiment_results}. 
  
\end{enumerate}

Regarding the details of the experiment participants, we recruited a group of sixty persons from the students of a software testing course: 49 of the participants successfully completed the experiment. There were nine females and 40 males, and the mean age was 23.6 years ($SD=1.1$). Two participants were left-handed, two participants wear glasses for both long and short distances, 15 wear glasses for long distances, 32 do not need prescription glasses, and only one participant changes between glasses for reading and glasses for looking at a distance. In their routine work (not in the experiments, where the environment was standardized), ten participants regularly use a touchpad as a primary pointing device, one uses a trackpoint and 39 use a mouse. 

The first group of 25 participants included three females and 22 males, and the mean age was 23.8 years ($SD=1.1$). One participant was left-handed. Eleven of the participants wear glasses for long distance vision in the first group. The second group of 24 participants included six females and 18 males, and the mean age was 23.5 years ($SD=1.2$). One participant was left-handed. Four participants wear glasses for long distance and two wear glasses for both long and short distance. The distribution of the pointing devices used in routine work was similar between groups. In each group, five participants use a trackpad and the others use a mouse.  

In the experiments, we took the following measures to prevent the impact of a possible learning effect: (1) participants started the experiment with a training scenario, and the results from these scenarios were not taken into account in the evaluation of the experimental data, and (2) we randomized the sequence of user scenarios to be executed by each of the participants and maintained an overall equal distribution of these sequences.

\subsection{User Scenarios in the Experiment}
\label{subsec:user_scenarios}

We considered the following user scenarios in the experiment:

\begin{enumerate}
\item Examine all photos from the given movie in the "Popular" section.

\item Count the number of movies in the category "TOP TV."

\item Check if there is a movie with given name in the category "TOP RATED."

\item Count the number of comedies in the category "TOP RATED." To determine if a movie is a comedy or not, use the movie metadata in its attributes.
  
\end{enumerate}

Scenario 1 was used as a training scenario to allow the experiment participants to become familiar with the testing environment. The results of this scenario were not evaluated further. Scenarios 2, 3 and 4 were used to collect data to adjust the model parametrization and the thresholds used in the UI verification rules.

\subsection{Implementation of the Proposed Automated UI Analysis and Testbed Setup}
\label{sec:implementation}

We implemented the proposed automated method in the development version of the used MBT platform\footnote{http://still.felk.cvut.cz/oxygen/}. In this environment, we created an abstract user scenario. We then mapped the steps of the abstract user scenario to the nodes $N$ and edges $E$ of the user interaction model $G$ to compose a tested user scenario $t$.

During the assessment of $t$ in $G$, the user scenario path $p(t)$ is found in $G$ and, subsequently, the values $nr(p(t))$ and $\mathcal{E} (p(t),C)$ are computed. The parametrization of context $C$ (configuration of the values of $UC$, $\mathcal{E}_{dev}$, $\mathcal{E}_{env}$ and $\delta$) is entered via two CSV files, to which a path is specified.

The computed results can be copied to the clipboard for further processing. For the experiment with the users (Step \ref{item.users_test} of the experiment method described in Section \ref{subsec:experiment_method}), the Smart TV environment web simulator with the testing app was deployed on a set of 20 workstations with the same hardware and operating system configuration to minimize possible bias caused by different hardware power or operating system configurations. 

The software simulation of the remote control device was available in the software runtime environment. To minimize the possible bias in results caused by different layouts or the simulated remote control, the same layout was configured for each of the participants.

\subsection{Experiment Results}
\label{subsec:experiment_results}

During the first phase of the experiment, we collected data from 25 user participants (group A) who successfully completed the assigned user scenarios. In total, the users completed 75 test scenarios and produced 75 user scenario paths. As mentioned previously, we did not consider the training scenario. Regarding the individual steps of the analyzed user scenario paths, we analyzed 2696 path steps, i.e., the transitions between nodes of the app model $N$ triggered by input actions invoked by pressing the UP, DOWN, LEFT, RIGHT, OK and BACK buttons on the remote control. To exclude excessively long transitions from the data processing, we set a threshold of 10 s. We considered user scenario path steps longer than 10 s to be biased, i.e., the user left the interaction with the UI for a certain time and then returned. In total, we excluded 49 steps, leaving 2647 steps in total to analyze. Table \ref{tab:group1_results_individual_buttons} presents a breakdown of the acquired data by individual input actions. In Table \ref{tab:group1_results_individual_buttons}, $valid$ stands for the number of valid actions, $invalid$ stands for the number of invalid actions, and SD stands for the standard deviation.

\begin{table}
\caption{Detailed results for individual input actions - experimental group A} 
\centering
\scriptsize
\begin{tabular}{|l|c|c|c|c|}
\hline
input action & avg. time [ms] & $valid$  & $invalid$ & avg. time SD \\ 

\hline
LEFT   & 1063 & 83 & 5 & 873.98  \\   \hline                
RIGHT &	975 &	1596 &	8 &	814.58 \\ \hline
UP &	687 &	4 &	0 &	250.58 \\ \hline
DOWN &	1173 &	36 & 	0 &	1383.17 \\ \hline
OK &	2418 &	508 &	31 &	1739.04 \\ \hline
BACK &	1293 &	420 & 	5 &	861.19 \\ \hline

\end{tabular}
  \label{tab:group1_results_individual_buttons}
\end{table}

Table \ref{tab:group1_results_scenarios} compares the results from the experiment with results obtained from the automated UI verification conducted on the MBT platform for the initial parametrization of the app model \ref{tab:initial_parametrization}. Scenario IDs (ID of $t$ in Table \ref{tab:group1_results_scenarios}) refer to the numbering in Section \ref{subsec:user_scenarios}. Scenario 1 was used as a training scenario and is not included in the analysis. The left part of the table presents the results from the experiment with the group of users. Here, $avg\_time$ means the average time needed to execute the scenario in milliseconds. SD stands for the standard deviation, and $avg\_stp$ represents the average number of steps in the user scenario paths executed by the experiment participants. The middle part of Table \ref{tab:group1_results_scenarios} presents the results of the analysis for the scenarios performed using the verification rules proposed in this paper and implemented in the MBT platform. Here, we present $|p(t)|$ and $\mathcal{E} (p(t),C)$ in milliseconds for each $t$. The right part of Table \ref{tab:group1_results_scenarios} compares the experimental results of the user group with the results obtained from the analysis conducted by the proposed method. Here, $DIFF_{stp} = \frac{|p(t)|}{avg\_stp} \cdot 100\%$ and $DIFF_{time} = \frac{\mathcal{E} (p(t),C)}{avg\_time} \cdot 100\%$.

\begin{table*}
\caption{Results for tested user scenarios - experimental group A} 
\centering
\scriptsize
\begin{tabular}{|l|c|c|c|c|c|c|c|c|}
\hline
\multirow{2}{*}{ID of $t$}
 & \multicolumn{4}{c|}{Experiment with users}
 & \multicolumn{2}{c|}{Proposed method} & 
 \multicolumn{2}{c|}{Differences}\\  \cline{2-9}

 & $avg\_time$ [ms] & $avg\_time$ SD & $avg\_stp$ & $avg\_stp$ SD & $\mathcal{E} (p(t),C)$  &  $|p(t)|$ & $DIFF_{time}$ & $DIFF_{stp}$ \\  \hline

2 &	28090 &	15310.52 & 27 &  15.89 & 20100 & 23 & 71.56\% & 85.19\% \\ \hline
3 &	19164 &	8095.88 & 16 & 7.34 & 7300 & 7 & 38.09\% & 43.75\% \\ \hline
4 &	91527 &	33128.95 & 61 & 18.67 & 97000 & 60 & 105.98\% & 98.36\% 
\\ \hline

\end{tabular}
  \label{tab:group1_results_scenarios}
\end{table*}

The relatively low correlation between $DIFF_{stp}$ and $DIFF_{time}$ indicates the suboptimality of the initial parametrization of the user interaction model. Additionally, $DIFF_{time}$ for scenario 4 indicates incorrect settings. For fewer steps (60 in the shortest path versus an average of 61 for the experiment participants), the $\mathcal{E} (p(t),C)$ computed by the proposed method is higher than $avg\_time$. Clearly, an update of the user interaction model parametrization is needed in the second iteration of the experiment. We further analyze and discuss the results, including the differences between the data obtained from the experiment with the users and the data obtained from the proposed automated analysis ($DIFF_{stp}$, $DIFF_{time}$ and their correlation), in Section \ref{subsec:discussion}.

In the second iteration of the experiments, we collected data from another 24 user participants (group B) that successfully completed the assigned user scenarios. Group~B was disjunctive from group A, and the participants were distributed randomly between the groups. In total, the users completed 72 test scenarios, producing 72 user scenario paths. Regarding the individual steps of the analyzed user scenario paths, we analyzed 2645 user scenario path steps. We kept the same threshold of 10 s to exclude excessively long transitions. In total, we excluded 59 steps, resulting in 2586 steps in total to analyze. Table \ref{tab:group2_results_individual_buttons} presents a breakdown of the acquired data by individual input actions for this second iteration of the experiment. After the analysis of the data presented in Tables \ref{tab:group1_results_individual_buttons}, \ref{tab:group1_results_scenarios}, \ref{tab:group2_results_individual_buttons}, and \ref{tab:group2_results_scenarios}, we adjusted the model parametrization. The adjusted values are presented in Table \ref{tab:adjusted_parametrization}.

\begin{table}
\caption{Detailed results for individual input actions - experimental group B} 
\centering
\scriptsize
\begin{tabular}{|l|c|c|c|c|}
\hline
input action & avg. time [ms] & $valid$  & $invalid$ & avg. time SD \\ 

\hline

LEFT&1178&	93&	3&	1027.16\\ \hline
RIGHT&	971&	1540&	11&	745.51\\ \hline
UP&	1251&	5&	0&	1196.70\\ \hline
DOWN&	1335&	30&	0&	1525.18\\ \hline
OK&	2179&	482&	44&	1459.61\\ \hline
BACK&	1243&	436&	1&	709.43\\ \hline

\end{tabular}
  \label{tab:group2_results_individual_buttons}
\end{table}

\begin{table}
\caption{Adjusted model parametrization} 
\centering
\scriptsize
\begin{tabular}{|l|c|c|c|c|}
\hline
action & \multicolumn{1}{l|}{$\delta(a(e))$} & \multicolumn{1}{l|}{$UC(a(e),C)$} & \multicolumn{1}{l|}{$\mathcal{E}_{dev}(C)$} & \multicolumn{1}{l|}{$\mathcal{E}_{env}(C)$} \\ \hline
LEFT   & 1000                              & 1.0                              & \multirow{6}{*}{1.0}               & \multirow{6}{*}{1.0}                    \\ \cline{1-3}
RIGHT  & 1000                              & 1.0                              &                                    &                                         \\ \cline{1-3}
UP     & 1000                              & 1.0                              &                                    &                                         \\ \cline{1-3}
DOWN   & 1250                              & 1.0                              &                                    &                                         \\ \cline{1-3}
OK     & 2000                              & 1.0                              &                                    &                                         \\ \cline{1-3}
BACK   & 1225                              & 1.0                              &                                    &                                         \\ \hline
\end{tabular}
  \label{tab:adjusted_parametrization}
\end{table}

Based on the above analysis, we also updated the values of the thresholds for the UI verification rules presented in Section \ref{sec:ui_analysis}. The value of $nr_{threshold}$ was kept at 1.5, $|p(t)|_{threshold}$ was set to 100 and $\mathcal{E}_{threshold}$ was set to 100000. Table \ref{tab:group2_results_scenarios} compares the results from the experiment with both groups with the results obtained from the automated UI verification performed on the MBT platform using the adjusted model parametrization (refer to Table \ref{tab:adjusted_parametrization}).

\begin{table*}
\caption{Results for tested user scenarios - experimental groups A and B} 
\centering
\scriptsize
\begin{tabular}{|l|c|c|c|c|c|c|c|c|}
\hline
\multirow{2}{*}{ID of $t$}
 & \multicolumn{4}{c|}{Experiment with users}
 & \multicolumn{2}{c|}{Proposed method} & 
 \multicolumn{2}{c|}{Differences}\\  \cline{2-9}

 & $avg\_time$ [ms] & $avg\_time$ SD & $avg\_stp$ & $avg\_stp$ SD & $\mathcal{E} (p(t),C)$  &  $|p(t)|$ & $DIFF_{time}$ & $DIFF_{stp}$ \\  \hline

\multicolumn{9}{|l|}{Experimental group A} \\  \hline

2 &	28090 &	15310.52 & 27 &  15.89 & 24250 & 23 & 86.33\% & 85.19\% \\ \hline
3 &	19164 &	8095.88 & 16 & 7.34 & 8000 & 7 & 41.74\% & 43.75\% \\ \hline
4 &	91527 &	33128.95 & 61 & 18.67 & 85275 & 60 & 93.17\% & 98.36\% 
\\ \hline

\multicolumn{9}{|l|}{Experimental group B} \\  \hline

2 &	32102 &	17728.78 & 30 &  12.19 & 24250 & 23 & 75.54\% & 76.67\% \\ \hline
3 &	17937 &	9463.59 & 16 & 7.38 & 8000 & 7 & 44.60\% & 43.75\% \\ \hline
4 &	85546 &	20019.32 & 60 & 14.31 & 85275 & 60 & 99.68\% & 100.00\% \\ \hline

\end{tabular}
  \label{tab:group2_results_scenarios}
\end{table*}

After the adjustment of the user interaction model parametrization, $DIFF_{time}$ improved compared to the results from the first iteration of the experiment (refer to Table \ref{tab:group1_results_scenarios}). The improvement in $DIFF_{stp}$ is not relevant to evaluate, as we compare the shortest path computed by the proposed method in the MBT environment with the paths taken in the app's UI by the participants from both experimental groups. However, the correlation between $DIFF_{time}$ and $DIFF_{stp}$ is more evident in this phase of the experiment, which indicates that the user interaction model parametrization was adjusted to be more accurate. We analyze and discuss the results further in Section \ref{subsec:discussion}.

\subsection{Discussion}
\label{subsec:discussion}

In this section, we discuss and analyze the experimental results in more depth. The first point to analyze is the significant difference between the results achieved by the experimental participants and the results produced by the proposed automated analysis in the case of scenario 3 (see Table \ref{tab:group2_results_scenarios}). The average length of the user path in this scenario was 16 steps for both experimental groups, whereas the result achieved by the automated analysis was 7. The rationale behind this difference is that there were two possible ways to iterate the list of movies in the tested app, either from an initial position in the list to the right or from the initial position to the left. Most users intuitively started iterating the list to the right, which required more steps. The proposed method took the shortest path to accomplish the task by iterating the list to the left. For both experimental groups, only a few actions with the UP button were performed (see Tables \ref{tab:group1_results_individual_buttons} and \ref{tab:group2_results_individual_buttons}) due to the nature of the tested user scenarios – the UP button was not practically required to accomplish the task. Due to low the frequency of UP actions, these data are not considered in the experimental evaluation.

For the RIGHT button, the average times of actions in Tables \ref{tab:group1_results_individual_buttons} and \ref{tab:group2_results_individual_buttons} are lower than those for the other buttons. The rationale behind this situation is that the RIGHT button was used in iterating the lists of the movies, which takes less time than needed to control other elements in the tested app. On the other hand, the average time for operations using the OK button is longer than that of the other buttons, which is expected because after pressing the OK button, the user usually moves to a new screen of the UI, where it takes time to choose the next action.

When analyzing the difference between the results of the experiment with the users and the results produced by the proposed automated analysis ($DIFF_{time}$ and $DIFF_{stp}$ in Tables \ref{tab:group1_results_scenarios} and \ref{tab:group2_results_scenarios}), in both phases of the experiment, the difference itself is not the primary indicator. As the proposed automated analysis can find a better path through the UI, the difference will be present in the comparison. In fact, what is important is the correlation between the $DIFF_{time}$ and $DIFF_{stp}$. In the first phase of the experiment (Table \ref{tab:group1_results_scenarios}), this correlation is present but not so strong. However, this correlation significantly increases in the second phase of the experiment (Table \ref{tab:group2_results_scenarios}), which indicates improvement in the configuration of the user interaction model.

Another point to discuss is the values of the thresholds for the UI verification rules presented in Section \ref{sec:ui_analysis}, namely, $nr_{threshold}$, $|p(t)|_{threshold}$ and $\mathcal{E}_{threshold}$. These thresholds can be set by the user based on judgment and experience with the developed app. However, some recommendations must be provided to potential users of the method. The results from the experiments showed that for Smart TV app, the values of the thresholds (especially $|p(t)|_{threshold}$) are significantly higher than the intuitively expected values for web, desktop or mobile app UI design. This difference can be explained by the relatively simple remote control device, which requires more user actions to reach particular elements of the UI, compare to web apps, for instance.

To compare the proposed approach with the available methods in the related areas, we start with manual usability testing of Smart TV apps. Compared to the proposed approach, which is automated, manual assessment of the usability of a smart TV app (e.g. \cite{shin2013smart, ingrosso2015ux, alam2017review}) might take more time and resources. From the performance viewpoint, the proposed automated approach makes the assessment process faster and hence, more repeatable after various changes in the developed smart TV app which might impact overall app development economics.

The proposed approach also differs from the previous proposals in the area of the automated usability testing of Smart TV apps based on a model created from the SUT UI. A comparable candidate here is an HSTM model by Cui et al. \cite{Cui2017}, based on a state machine and hierarchical structure of the app. In contrast to the approach proposed in this study, the HSTM based model is constructed by scanning the source code of the Smart TV app, whereas in our approach, UI screens are analyzed only. This fact might not impact the performance of the method itself; as it can be considered rather as an organizational constraint.

Additionally, an approach by Cui et al. is considered to be exhaustive as it will detect all the elements, including those that are not clickable. Hence, extensive filtering of those elements in the model is needed to generate an effective interaction model of the UI of the Smart TV app. This represents an extra step in the process and might impact its performance. Our approach does not require the source code of the app. Instead, we use a unique crawler to analyze the app's UI and detect the actual clickable elements, which, in turn, makes the proposed method more flexible. 

To compare the crawler proposed in this study to the alternatives available in the literature, crawlers that can be found, focus on different domains than the Smart TV apps (e.g. web applications \cite{Mesbah2012crawling}, thick-client app UIs \cite{memon2003gui} or mobile apps \cite{amalfitano2011gui,Amalfitano2015MobiGuitar,wang2014automatic}). Hence, they do not support the goals of our study, and it is a difficult task to compare their performance, as the domain of their operation differs significantly. 

Regarding the models used by mentioned crawlers, differences are also present in comparison to our study. Nguyen \textit{et al.} \cite{Nguyen2014:Guitar} are using an event-flow graph (EFG) as a model of the UI. Amalfitano \textit{et al.} \cite{Amalfitano2015MobiGuitar} employ a state machine as a model. These approaches are not applicable in the case of smart TV apps, which is caused by the different nature of the user's interaction with the app. To give an example, in a mobile app, the spatial distance between two icons (represented by two model states) is not relevant to the transition. In a smart TV app, this distance significantly matters and is expressed by a set of transitions. This difference leads to a different nature of the user interaction model.

Finally, regarding the model-checking approach for UI usability testing, the proposal of Harrison et al. \cite{harrison2017verification} can be compared with our approach. However, Harrison et al. are using temporal logic as an underlying model of the SUT and focuses on the verification of the UI of medical devices. In comparison, our approach uses the User Interaction Model based on a directed graph, and the primary goal is to identify the possible UI design sub-optimalities, so these two approaches are conceptually similar only. 

To conclude, we have identified no direct alternative to the approach proposed in this study and from this viewpoint, the proposed method is an original contribution to the filed of automated UI testing of Smart TV apps, based on automated creation of the User Interaction Model from the app UI.

\section{Threats to Validity}
\label{chap:threats_to_validity}

In this section, we discuss issues that might affect the accuracy or objectivity of the results. For each issue, we also consider its possible impact and the actions we took to mitigate the impact.

The first possible concern is that the learning effect during the experiments might bias the results. We prevented the learning effect by two measures: (1) each participant started the interaction with the UI with a training user scenario that was not included in the evaluation results, and (2) each participant was presented a randomized sequence of user scenarios. We kept the distribution of the sequences of user scenarios equal between the experimental groups.

Another concern that might be raised regarding the experiments is the simulation of the Smart TV environment that was used instead of a real Smart TV device, which might influence the measured data and, as a consequence, the accuracy of the suggested user interaction model parameterization (constants $UC$, $\mathcal{E}_{dev}$, $\mathcal{E}_{env}$ and $\delta$). However, the principle of the method and its use case is not affected by this possible limitation. For the Smart TV app, only the user interaction model parameterization constants have to be adjusted. Moreover, the same concern can be raised in the case of different types of remote controls. Here, the composition of the buttons on the remote control and general ergonomics of the device might influence the parametrization.

Additionally, the accuracy of $UC$, $\mathcal{E}_{dev}$, $\mathcal{E}_{env}$ and $\delta$ can be influenced by the set of user scenarios and analyzed apps. Achieving perfect and exact parametrization values is not a realistic task and is not a reasonable goal. The proposed method works within a certain tolerance given by the thresholds $nr_{threshold}$, $|p(t)|_{threshold}$ and $\mathcal{E}_{threshold}$, which can be adjusted by the users of the method.

\section{Conclusion}\label{conclusion}

The proposed combination of the SUT user interaction model reconstructed from an actual Smart TV application with a set of verification rules aimed to assess the feasibility and efficiency of user tasks (being the major quality characteristic in usability testing) is, to the best of our knowledge, an original attempt in the field, as such a study has not been published previously. 

In contrast to the manual usability testing techniques, the proposed method is automated; therefore, the proposed technique is faster and the possibility of human-made mistakes is lower. On the other hand, if the method is not configured correctly, the findings might be misleading. We minimized this effect by conducting experiments in which we adjusted values of the configuration constants $UC$, $\mathcal{E}_{dev}$, $\mathcal{E}_{env}$ and~$\delta$~based on the results of independent test with a group of users.

\bibliographystyle{IEEEtran}
\bibliography{sample-bibliography}

\begin{IEEEbiography}[{\includegraphics[width=1in,height=1.1in,clip,keepaspectratio]{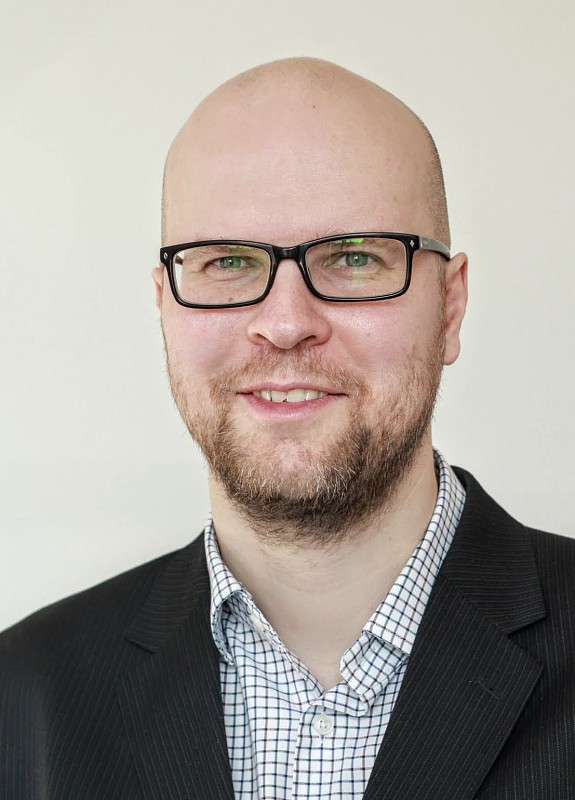}}]{Miroslav Bures}
 received his Ph.D. at Czech Technical University in Prague. His research interests are model-based testing, path-based testing, data consistency testing and combinatorial interaction testing, effective test automation (test automation architectures, assessment of automated testability, economic aspects) and quality assurance methods for Internet of things solutions.
 \end{IEEEbiography}

\begin{IEEEbiography}[{\includegraphics[width=1in,height=1.1in,clip,keepaspectratio]{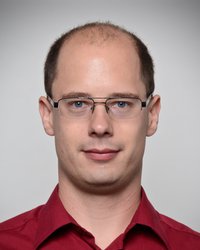}}]{Miroslav Macik} received his Ph.D. at Czech Technical University in Prague. He currently works at the same institution as a researcher in the field of Human-Computer Interaction. His research focuses on model-based design and evaluation, haptic interaction, accessibility. 
 \end{IEEEbiography}

 \begin{IEEEbiography}[{\includegraphics[width=1in,height=1.1in,clip,keepaspectratio]{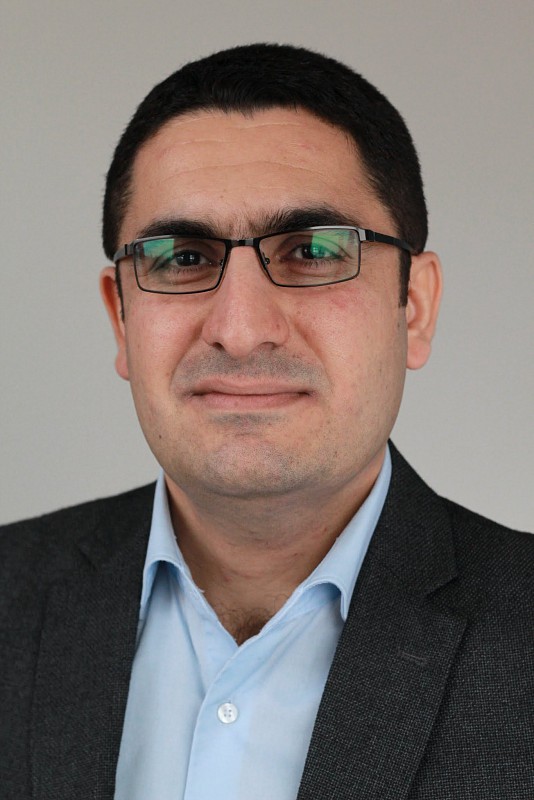}}]{Bestoun S. Ahmed}
 obtained his Ph.D. from University Sains Malaysia (USM) in 2012. Currently, he is a senior lecturer at the department of mathematics and computer science, Karlstad University, Sweden. His main research interest include Combinatorial Testing, Search Based Software Testing (SBST), and Applied Soft Computing.
\end{IEEEbiography}

\begin{IEEEbiography}[{\includegraphics[width=1in,height=1.1in,clip,keepaspectratio]{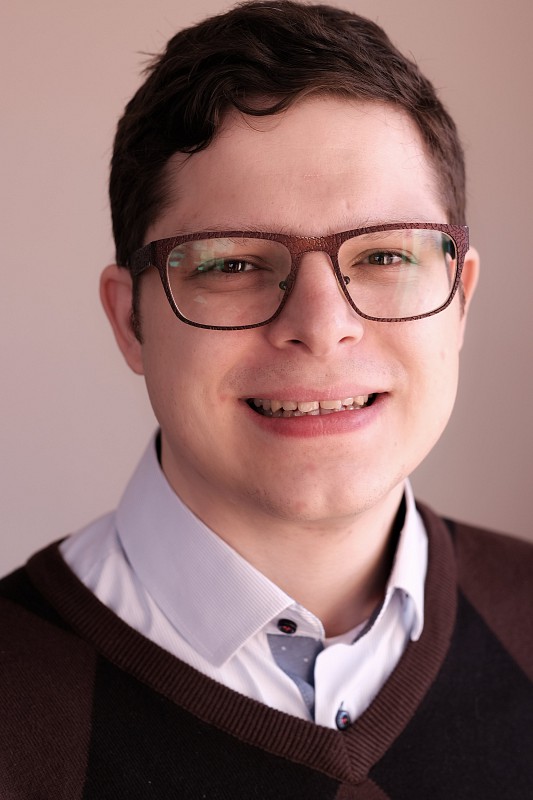}}]{Vaclav Rechtberger} is PhD. student in Software testing Intelligent Lab (STILL), dept. of Computer Science and Engineering, Czech technical University in Prague. His focus is Model-based Testing, test automation and testing of Internet of Things systems. \end{IEEEbiography}

\begin{IEEEbiography}[{\includegraphics[width=1in,height=1.1in,clip,keepaspectratio]{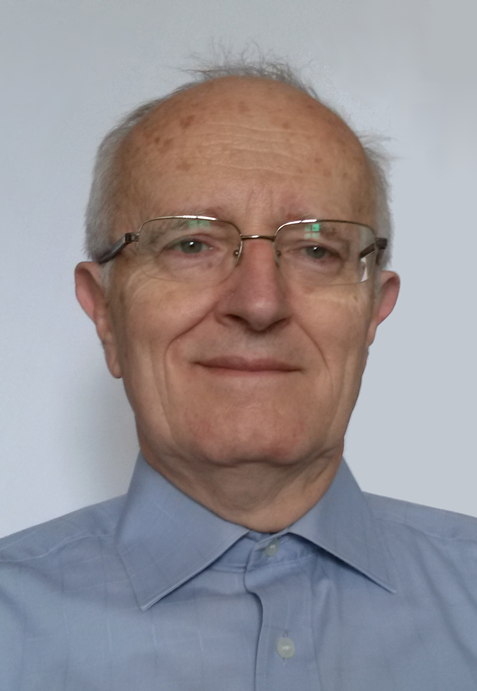}}]{Pavel Slavik} is full Professor of Computer Science and member of HCI group at Czech Technical University in Prague. His fields of interest are visualization, usability and accessibility. He  served in the past as an IPC member for several HCI conferences.\end{IEEEbiography}

\end{document}